\documentclass[doublecol,linenumbers]{epl2} 
\usepackage{amsmath}
\usepackage{amsfonts}
\usepackage{multicol}
\usepackage{caption}
\usepackage{epsfig}
\usepackage{url}
\usepackage{soul}
\setstcolor{red}
\usepackage{comment}
\title{Inferring Multi-Period Optimal Portfolios via Detrending Moving Average Cluster Entropy}
\shorttitle{Inferring Multi-Period Optimal Portfolios} %Insert here a short version of the title if it exceeds 70 characters
\author{P. Murialdo\inst{1} \and L. Ponta\inst{2} \and A. Carbone\inst{1}}
\shortauthor{P. Murialdo, L. Ponta,  A. Carbone}
\institute{                      \inst{1} Politecnico di Torino, corso Duca degli Abruzzi 24, 10129 Torino, Italy\\
  \inst{2} Università Cattaneo LIUC, Castellanza, Italy
}
\abstract{Despite half a century of research, there is still no general agreement about the optimal approach to build a robust multi-period portfolio. We address this question by  proposing the detrended cluster entropy approach to estimate the portfolio weights of high-frequency market indices. The information measure  produces
reliable estimates of the portfolio weights gathered from the
real-world market data at varying temporal horizons. The portfolio exhibits a high level of diversity, robustness and stability as it is not affected by the drawbacks of traditional mean-variance approaches.}
\begin{document}
\maketitle
\section{Introduction}
Markowitz mean-variance approach \cite{markowitz52portfolio} estimates minimum risk and maximum return for portfolio optimization models in financial decision-making processes. 
Variants of the original mean-variance approach have been proposed  integrating financial concepts and tools such as: capital asset pricing  \cite{evans1968diversification}, Sharpe ratio \cite{sharpe64capital},  excess growth rate  \cite{fernholz2002stochastic}  Gini-Simpson index \cite{woerheide1992index}.
However, contrary to the aim of diversifying, mean-variance approaches yield weights strongly concentrated on some assets, resulting in low diversity of the portfolio. 
\par
Traditional approaches have shown even more dramatic limits in multiple horizons and out-of-sample estimates, where nonstationarities and fat tails of the price return distribution come unavoidably into play \cite{Hakansson1971Multi,Gressis1976Multiperiod}. As high-frequency data become available, microstructure noise increasingly becomes dominant in the returns and volatility series, affecting portfolio performance by reducing signal-to-noise ratio, particularly over multiple periods.   Larger sampling intervals could reduce the effect of microstructure noise but with the evident disadvantage of not making full use of the available data.
Dynamic readjustment of the portfolio and sequential wealth re-allocation to selected assets in consecutive trading periods is a key requirement to gather relevant news. Regretfully, errors related to the microstructure noise and non-normality of financial series  are particularly relevant when portfolio weights should be estimated at different horizons
\cite{boyd2017multi,oprisor2021multi}. 
\par
Despite numerous and prominent efforts, the efficacy of quantitative methods of portfolio allocation still remains an open issue, leaving the financial community with the deceiving impression that the \textit{naive equally-weighted $1/{\cal{N}_{\cal{A}}}$ portfolio of ${\cal{N}_{\cal{A}}}$ assets} is not yet significantly outperformed by other approaches to portfolio optimization
\cite{demiguel2009optimal,fletcher2011optimal,frahm2013diversification}
\par
Over the past decades, a growing wave of interest has been directed towards the analysis  of nonlinear interactions arising in complex systems in different contexts. 
Complex systems exhibit remarkable features related to patterns  emerging  from the
seemingly random structure of time series due to the interplay of long- and short-range correlated processes
 \cite{raberto2001agent,di2003scaling,yamasaki2005scaling,yakovenko2009colloquium,chakraborti2011econophysics,carbone_tails_2007,kwapien2012physical,sornette2017stock}.
Hence, entropy and other information measures have increasingly found applications in complex systems science and, in particular, in economics and finance.  As a tool for quantifying \textit{dynamics}, entropy has been adopted for shedding light on fundamental aspects of asset pricing models \cite{backus2014sources,ghosh2017consumption}.
 As a tool for quantifying  \textit{diversity}, entropy has been exploited for mitigating drawbacks of  traditional portfolio strategies. In this specific context, entropy is considered a convenient instrument of shrinkage and
 dispersion for the traditional portfolio weights distribution lacking reasonable diversity degree  \cite{philippatos1972entropy,ou2005theory, bera2008optimal,ormos2014entropy,batra2020portfolio,lim2020portfolio,simonelli2005indeterminacy,zhou2013applications,zhou2013portfolio,meucci2009risk,meucci2016dynamic,kirchner2011measuring,vermorken2012diversification,yu2014diversified,pola2016entropy,contreras2017construction,bekiros2017information,chen2017study}. 
\par
The detrended cluster entropy approach \cite{carbone2004analysis,carbone2007scaling,carbone2013information} has been adopted  to investigate several assets  over a single period in \cite{ponta2018information}.  An information measure of diversity, the \textit{cluster entropy index} $I(n)$, was put forward by integrating the entropy function over the cluster dimension $\tau$, with the moving average window $n$ as a parameter. It was shown that the \textit{cluster entropy index} $I(n)$ of the volatility is significantly market dependent. Hence, the construction of an efficient single-period \textit{static} portfolio based on the \textit{cluster entropy index} $I(n)$  has been proposed and compared to traditional \textit{mean-variance} and \textit{equally-weighted $1/\cal{N}_{\cal{A}}$ portfolios of $\cal{N}_{\cal{A}}$ assets}. 
\par
 The cluster entropy approach was extended to multiple temporal horizons $\cal{M}$  showing an interesting and significant horizon dependence particularly in long-range correlated markets in \cite{ponta2021information,murialdo2020long}. Artificially generated price series have been systematically analysed in terms of the cluster entropy dependence on the horizons $\cal{M}$. The cluster entropy for Fractional Brownian Motion (FBM) and Generalized AutoRegressive Conditional Heteroskedasticity (GARCH) processes proved unable to reproduce real markets dynamics. Conversely,  for  Autoregressive Fractionally Integrated Moving Average (ARFIMA), the cluster entropy proved able to replicate asset behaviour. Hence, results obtained by the cluster entropy approach on real-world market series are consistent with the hypothesis of financial processes deviating from \textit{i.i.d.} stochastic processes, proving the ability of the detrended cluster entropy approach of capturing the important statistical features and stylized facts of the real world markets \cite{ponta2021information,murialdo2020long}.
\par
In this work, building on the method proposed in \cite{ponta2018information,ponta2021information} and  the   systematic analysis conducted in \cite{murialdo2020long},   we discuss how a robust multi-period dynamic portfolio can be constructed via the cluster entropy of \textit{return} and  \textit{volatility} of $\cal{N}_{\cal{A}}$ assets.  The cluster entropy index $I(n)$  will be \textit{dynamically} implemented  at  different temporal horizons . The  multi-period  portfolio is estimated over five assets (${\cal{N}_{\cal{A}}}= 5$) and  twelve consecutive monthly periods over one year (${\cal{M}}=\{0 ,\dots , 12\}$).
 %%%
\section{Mean-Variance Approach to Portfolio Construction}
\label{Sec:Portfolios}
A portfolio is a vector $w=\left\{w_{1}, w_{2}, w_{2}, \ldots, w_{{\cal{N}_{\cal{A}}}}\right\}$, satisfying $\sum_{i=1}^{\cal{N}_{\cal{A}}} w_{i}=1$, representing the relative allocation  of wealth in asset $i$. Let ${r}_{t,i}$ be the return of the asset $i$ at time $t$. The expected return of the portfolio $\mu({r}_{p})$ can be written in terms of the expected return $\mu({r}_i)$ of each asset as:
 \begin{equation}
 \label{eq:Portfoliomean}
\mu\left(r_{p}\right)=w_{1} \mu\left(r_{1}\right)+\cdots+w_{\cal{N}_{\cal{A}}} \mu\left(r_{{\cal{N}_{\cal{A}}}}\right)
\end{equation}
  Let $\sigma_{i}$ indicate the standard deviation of the return $r_{t,i}$ of the asset $i$, $\sigma_{i j}$ the covariance between $r_{t,i}$ and $r_{t,j}$. The variance of the portfolio return $\sigma^{2}\left(r_{p}\right)$  can be written as:
\begin{equation}
\label{eq:Portfoliovariance}
\begin{split}
\sigma^{2}\left(r_{p}\right) & = w_{1}^{2} \sigma_{1}^{2}+\cdots+w_{{\cal{N}_{\cal{A}}}}^{2} \sigma_{{\cal{N}_{\cal{A}}}}^{2}+ \sum_{i=1}^{{\cal{N}_{\cal{A}}}} \sum_{j=1 \atop i \neq j}^{{\cal{N}_{\cal{A}}}}w_{i} w_{j} \sigma_{i} \sigma_{j}\\
& = \sum_{i=1}^{\cal{N}_{\cal{A}}} w_{i}^{2} \sigma_{i}^{2}+\sum_{i=1}^{{\cal{N}_{\cal{A}}}} \sum_{j=1 \atop i \neq j}^{{\cal{N}_{\cal{A}}}} \sigma_{i j} w_{i} w_{j}
\end{split}
\end{equation}
According to the Markowitz portfolio strategy, the weights $w_{i}$, for $i=1,2,3 \ldots \cal{N}_{\cal{A}}, \quad$ are chosen to minimize the variance of the return  $\sigma^{2}\left(r_{p}\right)$ under the constraint of an expected portfolio return $\mu({r}_{p})$.
 The performance of the mean-variance portfolio can be maximized in terms of the Sharpe ratio $R_{\cal{S}}$  defined as:
\begin{equation}
\label{eq:Sharperatio}
R_{\cal{S}}=\frac{\mu\left(r_{p}\right)}{\sqrt{\sigma^{2}\left(r_{p}\right)}} \quad .
\end{equation}
The maximization of the Sharpe ratio, with  maximum portfolio return eq.~(\ref{eq:Portfoliomean}) and  minimum portfolio variance  eq.~(\ref{eq:Portfoliovariance}), is commonly adopted in the standard portfolio theory to nominally yield the optimal weights $w_i$.  
However, as eq.~(\ref{eq:Portfoliomean}) and  eq.~(\ref{eq:Portfoliovariance})  imply normally distributed stationary return series, the approach is flawed at its foundation. Thus, very biased portfolio weights are obtained as it could be reasonably expected with the asymmetric and heavy tailed return distributions of  real-world markets. Several variants of the original theory have been thus  proposed to overcome those limits. The poor accuracy and lack of diversity in the estimation of portfolio weights, with the  transaction costs involved in the  optimization constraints, have been soon recognized as limits of the applicability of mean-variance based models \cite{demiguel2009optimal,fletcher2011optimal,frahm2013diversification}.
\par
\begin{figure*}[h]
 \centering
 \includegraphics[width=0.32 \textwidth]{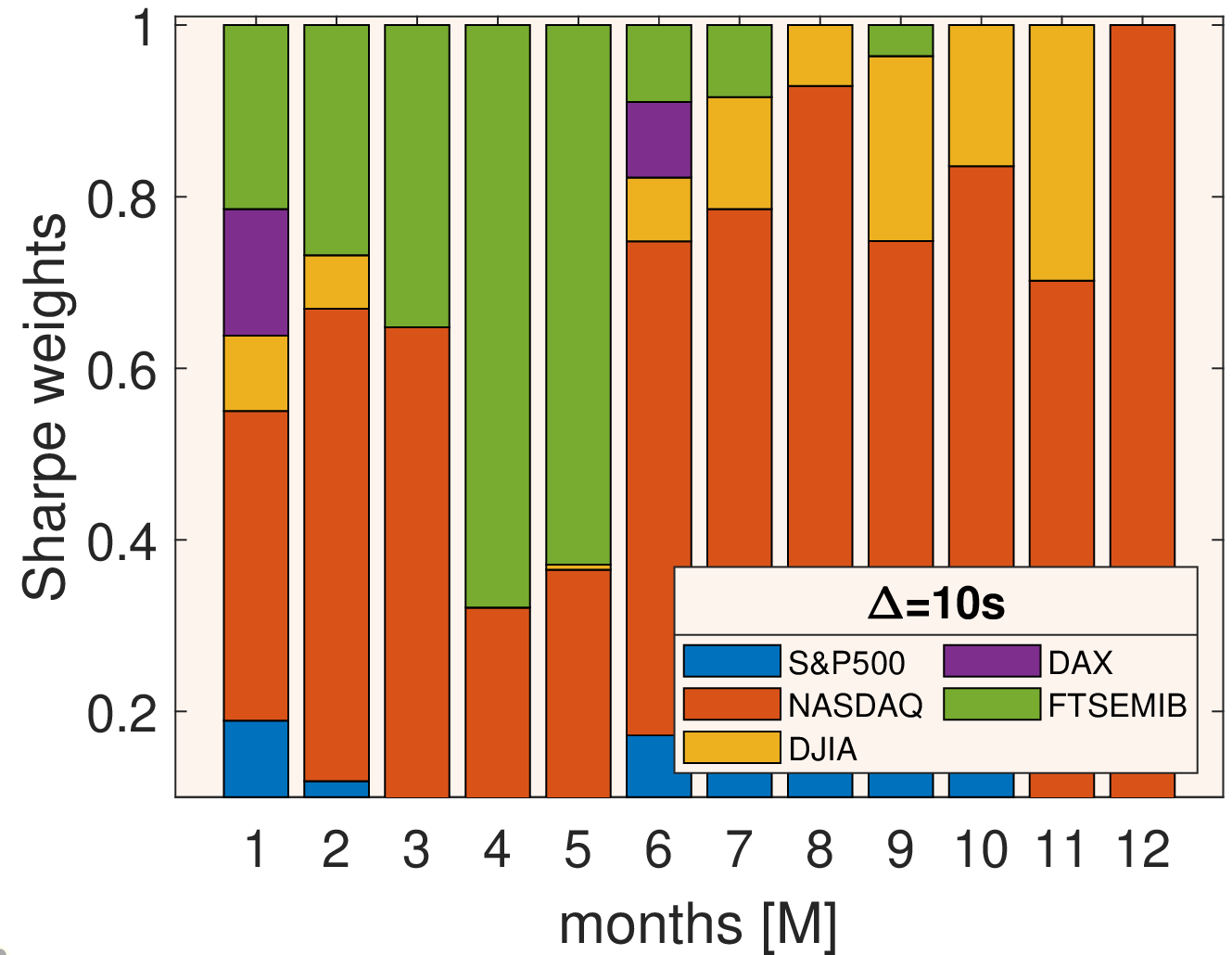}
 \includegraphics[width=0.32 \textwidth]{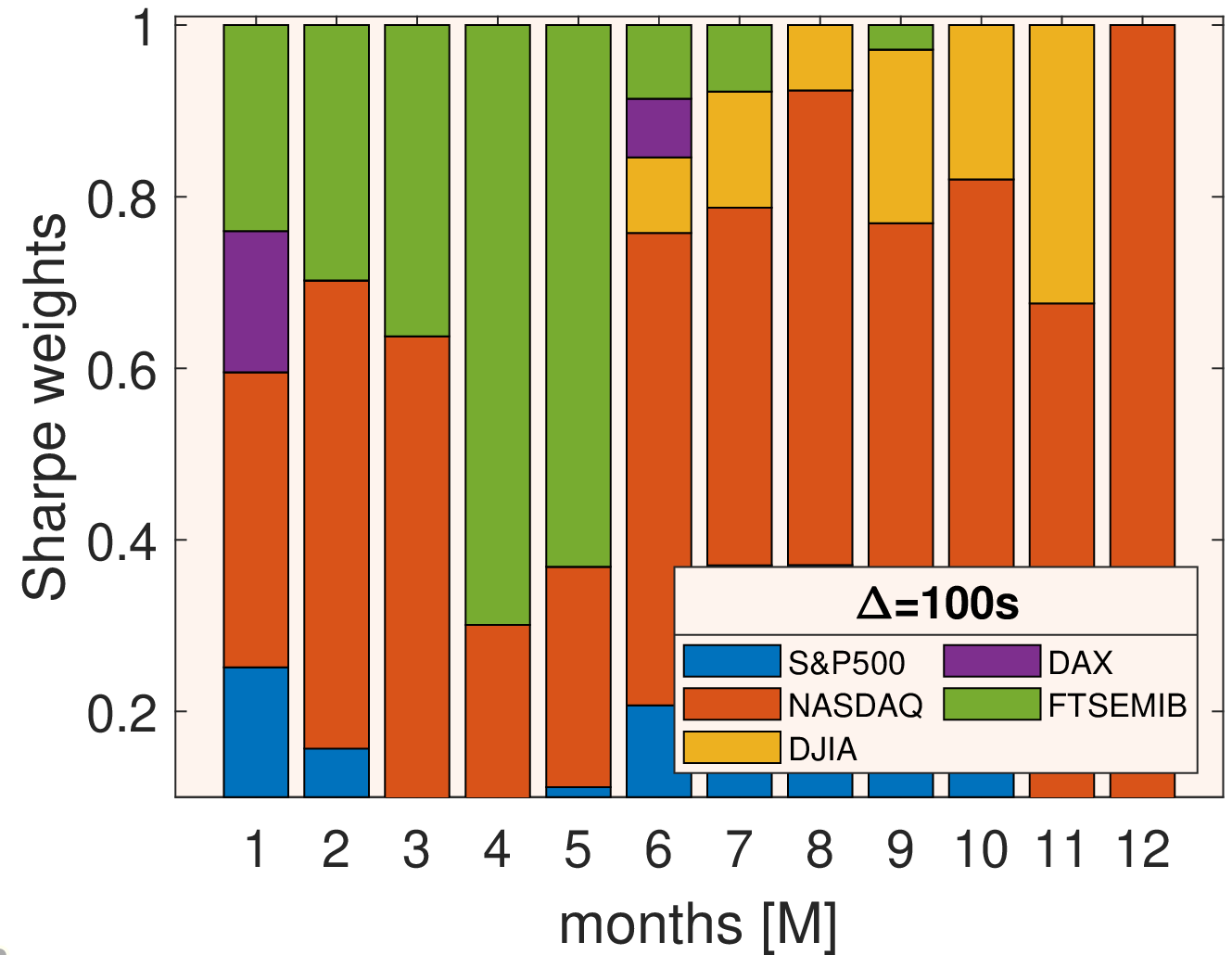}
  \includegraphics[width=0.32 \textwidth]{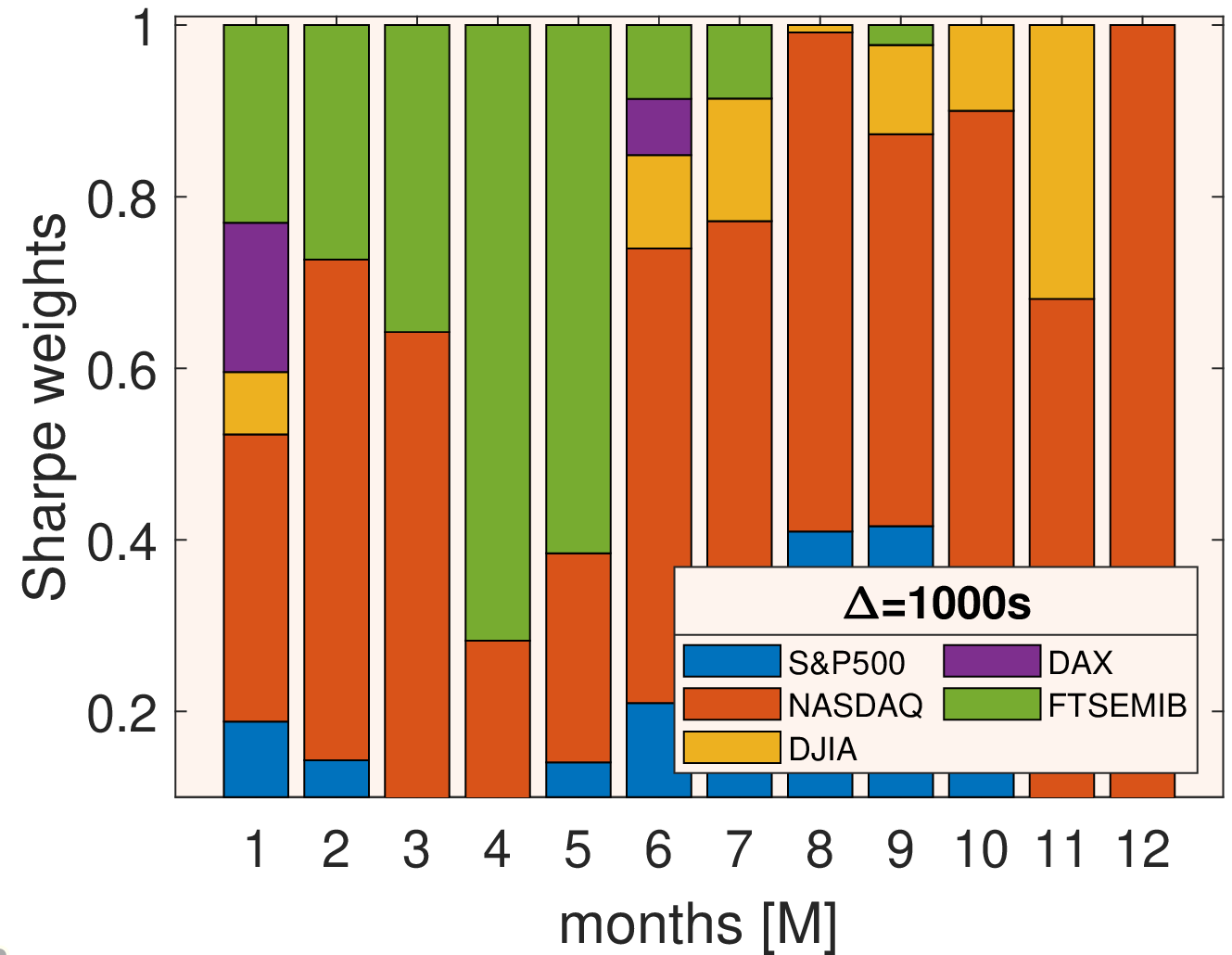}
 \caption{\label{fig:barsharpeWeights} Portfolio weights $w_i$ vs. investment horizon $\cal{M}$. The weights are obtained by using the standard mean-variance estimates eqs.~(\ref{eq:Portfoliomean},\ref{eq:Portfoliovariance})  which maximize the Sharpe ratio $R_{\cal{S}}$ eq.~(\ref{eq:Sharperatio}). High-frequency data for the assets described in Table \ref{tab:data} have been used for the estimates. Raw data prices have been downloaded from the Bloomberg terminal. The data are sampled to obtain equally spaced series with equal lengths. Sampling interval is indicated by $\Delta$. The main drawbacks of the traditional portfolio strategy as a results of the non-normality and non-stationarity of real-world asset price distributions can be easily noted: (\textit{i}) weights are concentrated towards the extremes  (i.e. 0 and 1 values are very likely, thus contradicting the principle of high-diversified portfolios); (\textit{ii}) weights exhibit abrupt changes over consecutive horizons  (Further results of portfolio weights can be found in the figures included in the Supplementary Material attached to this letter).
}
 \end{figure*}

\begin{figure*}[h]
 \centering
 \includegraphics[width=0.32 \textwidth]{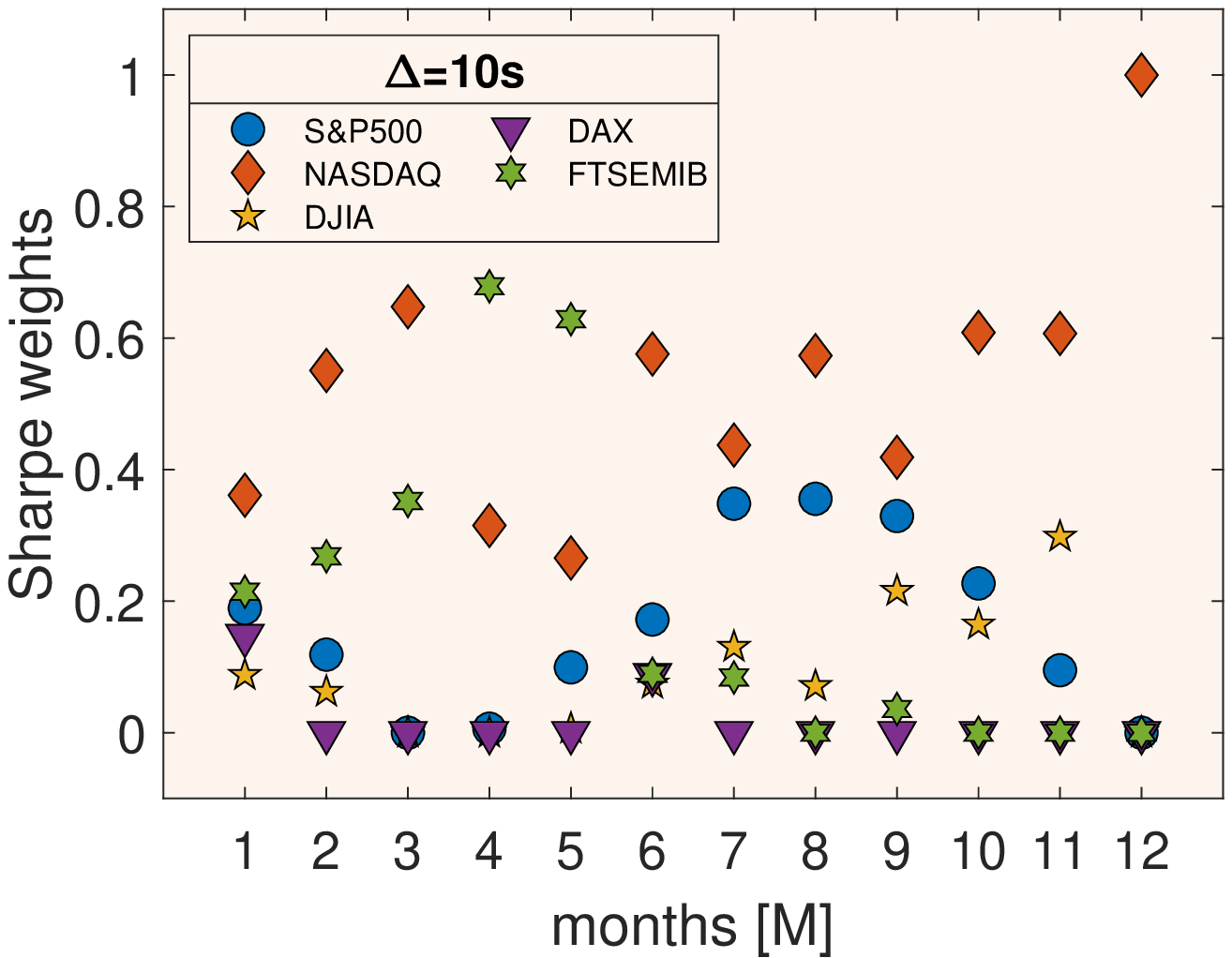}
 \includegraphics[width=0.32 \textwidth]{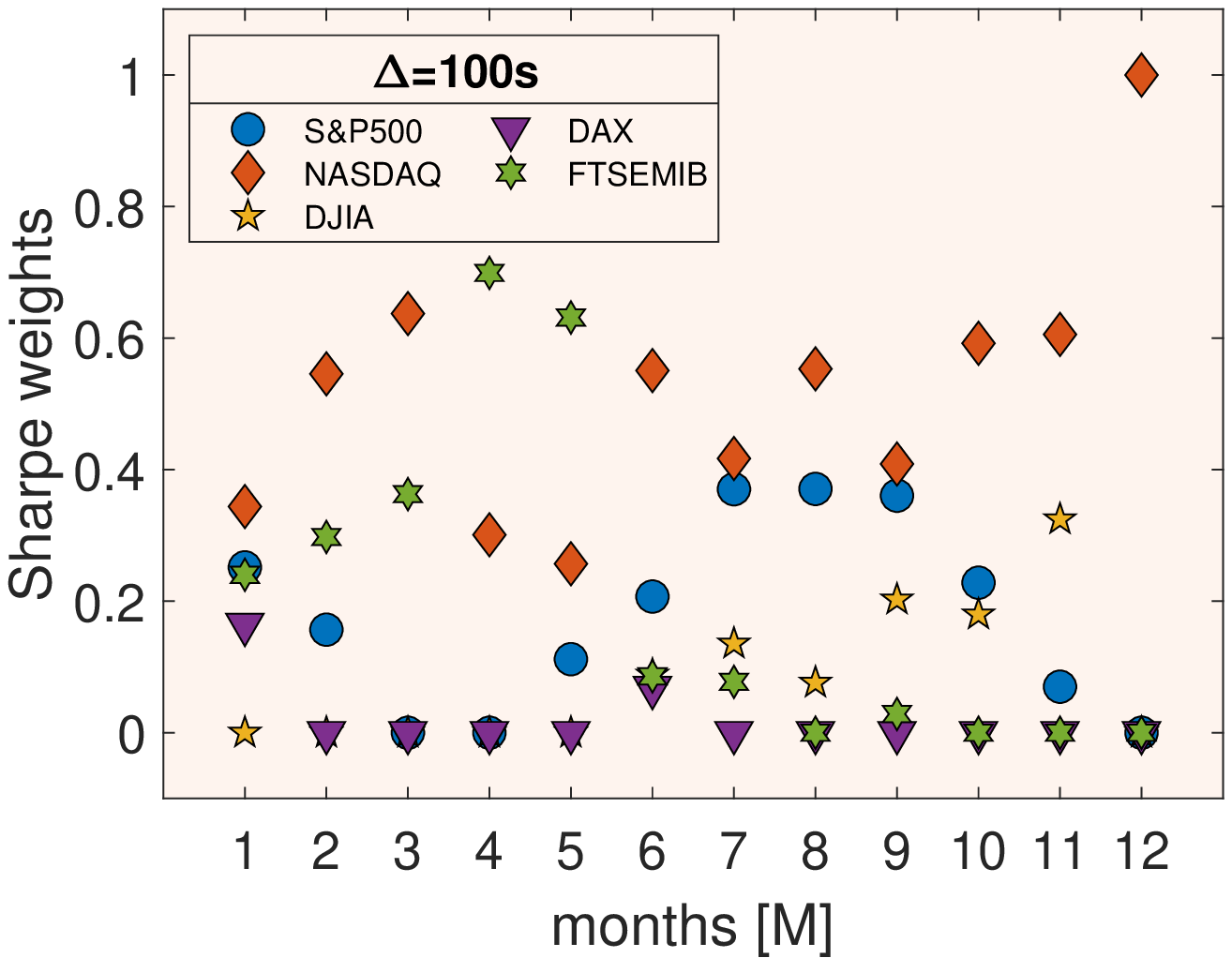}
 \includegraphics[width=0.32 \textwidth]{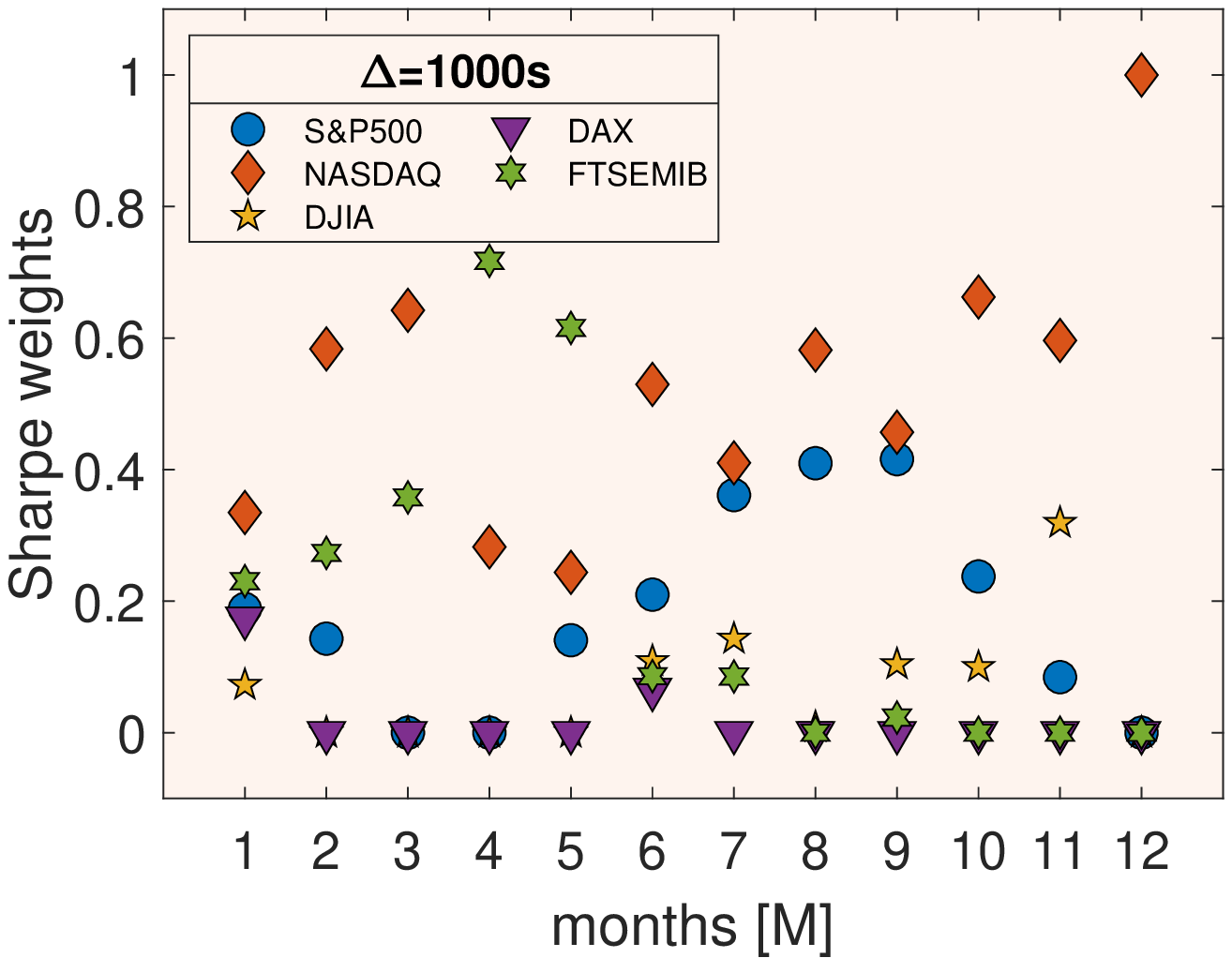}\\
 \caption{\label{fig:sharpeWeights} Same as in fig. 1  but with scattered plots.
}
 \end{figure*}

 
To fully appreciate the errors in the weights yielded by the traditional mean-variance approach, the Sharpe-ratio  has been estimated on tick-by-tick data of the high-frequency markets described in Table \ref{tab:data} by using the code provided by the MATLAB Financial Toolbox.
%%%%%%%%%%%%  TABLE %%%%%%%
\begin{table*}[h]
\caption{\textit{Asset Description}. Assets and  metadata are downloaded from Bloomberg terminal. Tick duration (time interval between individual transactions) is of the order of second for all the markets.
The S\&P 500 index includes 500 leading companies, but two types of shares are mentioned for 5 companies thus 505 assets are to be considered for calculations. Analogously, for NASDAQ index the number of members is 2570, for DJIA index is 30, for DAX index is 30 and for FTSEMIB is 40. See Bloomberg website for further details on index composition.
}
\centering
\label{tab:data}
\begin{tabular}{llcrrr}
{Ticker} & {Name} & { Country} & {Currency} & {Members}  & {Length}  \\
\hline
    NASDAQ & Nasdaq Composite & US & USD & 2570 & 6982017 \\
    S\&P500 & Standard \& Poor 500  & US & USD & 505 & 6142443 \\
    DJIA & Dow Jones Ind. Avg & US & USD & 30 & 5749145 \\
    DAX & Deutscher Aktienindex & DE & Euro & 30 & 7859601 \\
    FTSEMIB & Milano Indice di Borsa & UK & Euro & 40 & 11088322 \\
\end{tabular}
\end{table*}
The values of portfolio weights, which maximize the Sharpe ratio, are shown in fig.~\ref{fig:barsharpeWeights} and fig.~\ref{fig:sharpeWeights}.  Raw market data are sampled to yield equally spaced series with equal lengths. Sampling intervals are indicated by $\Delta$. The Sharpe ratio maximization is performed on twelve multiple horizons $\cal{M}$. One can note (i) the unreasonably high variability of the weights of the same assets  over  consecutive periods and (ii) the biased distribution of the portfolio weights oriented towards the riskiest assets rather than a diversified portfolio, resulting in a quite scary and disappointing overall investment scenario.

\section{Cluster Entropy Approach to Portfolio Construction}
Among several alternatives proposed to build an effective portfolio, entropy-based tools have been recently developed, supported by the general idea that entropy itself is a measure of diversity.
\par
 Entropy-based portfolio inference is based on Shannon entropy, defined as 
\begin{equation}
\label{eq:Shannonentropy}
S(P_i)=-\sum_{i} P_{i} \ln P_{i},
\end{equation}
%%%%%%%%%%%%%%%%
$P_{i}$ being the probability associated to a given stochastic variable relevant to the asset $i$.
\par
Initial attempts have introduced the portfolio weights, obtained by Markowitz based approaches, into eq.~(\ref{eq:Shannonentropy}). Portfolio weights $w=\left(w_{1}, w_{2}, \ldots, w_{{\cal{N}_{\cal{A}}}}\right)^{\prime}$ of ${\cal{N}_{\cal{A}}}$ risky assets, with $w_{i} \geq 0, i=1,2, \ldots, {\cal{N}_{\cal{A}}}$ and $\sum_{i=1}^{{\cal{N}_{\cal{A}}}} w_{i}=1,$ have the structure of a probability distribution, thus Shannon entropy can be written as:
\begin{equation}
\label{eq:ShannonWeight}
S(w_i)=-\sum_{i=1}^{\cal{N}_{\cal{A}}} w_{i} \ln w_{i}
\end{equation}
 One can immediately note that with \textit{equally distributed naive weights} $u_{i}=1/\cal{N}_{\cal{A}}$ for all $i$, $S(w_i)$ reaches its maximum value $\ln \cal{N}_{\cal{A}}$. Conversely, when $w_{i}=1$ for the asset $i$ and $w_{i}=0$ for the others, $S(w_{i})=0$.
 \par
 Portfolio optimization, based on Kullback-Leibler minimum cross-entropy principle, was proposed in \cite{bera2008optimal}:
 \begin{equation}
\label{eq:Kullbackweights}
S(w_i,u_i)=-\sum_{i=1}^{\cal{N}_{\cal{A}}} w_{i} \ln \frac{w_{i}}{u_{i}},
\end{equation}
   $S(w_i,u_i)$ is minimized with respect to the reference  distribution $u_i$.  If the equally distributed probability of the weights  $u_{i}=1/\cal{N}_{\cal{A}}$ is taken as reference, the approach provides a shrinkage of the poorly diversified weights towards the uniform distribution.
\par
When entropy approaches are used, the portfolio is generally shrinked toward an equally weighted portfolio, which corresponds to the maximally diversified portfolio yielded by the  \textit{equally distributed naive weights} $1/{\cal{N}_{\cal{A}}}$ rule.
However, the portfolios yielded by either  eq.~(\ref{eq:ShannonWeight}) or eq.~(\ref{eq:Kullbackweights})  are still affected by the native limitation of operating with the weights $w_{i}$ estimated by the traditional approach, requiring normally and stationary distributed data. Thus, very critical performances are obtained when multiple horizons and out-of-sample estimates are considered. 
\par
The \textit{Detrending Moving Average (DMA) cluster entropy method} goes beyond this limit as it does not rely on the assumption of Gaussian distributed returns.  
The  \textit{DMA cluster entropy} approach to portfolio optimization   relies on the general Shannon functional eq.~(\ref{eq:Shannonentropy}). The probability distribution function of each asset $i$, defined as $P_i(\tau_j,n)$, is
obtained by intersecting the asset time series $y(t)$ with its moving average series $\tilde{y}_n(t)$ \cite{carbone2004analysis,carbone2007scaling}. 
The simplest moving average is defined at each $t$ as the sum of the $n$ past observations from $t$ to $t-n+1$, namely $\tilde{y}_n(t) = \,{1}/{n} \sum_{k = 0}^{n - 1} y(t-k)$. However, more general detrending moving average cluster distributions can be generated with higher-order moving average polynomials as shown in \cite{arianos2011self,carbone2016detrending}.
\par 
Consecutive intersections of the time series and their moving averages yield several sets  of \textit{clusters}, defined as the portion of the time series between two consecutive intersections of $y(t)$ with  $\tilde{y}_n(t)$. The cluster duration is equal to:
$\tau_j \equiv || t_j -t_{j-1} ||$
where $t_{j-1}$ and $t_j$ refers to consecutive intersections of $y(t)$ and $\tilde{y}_n(t)$. 
For each moving average window $n$, the probability distribution function $P_i(\tau_j, n)$, i.e. the frequency of the  cluster  lengths $\tau_j$, can be obtained by counting the number of clusters $\mathcal{N}_{\cal{C}}(\tau_j,n)$ with length $\tau_j$, $j \in \{1,N-n-1\}$. The probability distribution function $P_i(\tau_j,n)$ results:
\begin{equation} \label{eq:probdistr}
P_i(\tau_j,n) \sim \tau_j^{-D} \mathcal{F}(\tau_j,n) \quad,
\end{equation}
with $D = 2-H $ the fractal dimension 
and $H$ the Hurst exponent of the series ($0 < H < 1$), according to the widely accepted framework of power-law scaling of temporal correlation (see e.g. \cite{carbone2003scaling,ponta2021information,murialdo2020long,rak2015detrended}).
The term $\mathcal{F}(\tau_j,n)$ in eq. (\ref{eq:probdistr}) takes the form:
\begin{equation}\mathcal{F}(\tau_j,n) \equiv e^{-\tau_j/n}\quad,
\end{equation}
to account for the drop-off of the power-law behavior and the onset of the exponential decay when $\tau \geq n$ due to the finiteness of $n$. When $n \rightarrow 1$, a set of the order of $N$ clusters with lengths  centered around a single $\tau$ value. When $n \rightarrow N$, that is when $n$ tends to the length of the whole sequence, only one cluster with $\tau = N$ is generated.  Intermediate values of $n$ produces the broad distribution of cluster durations.
\par
When the probability distribution eq.~(\ref{eq:probdistr}) is fed into the Shannon functional eq.~(\ref{eq:Shannonentropy}),  the entropy $S_i(\tau_j,n)$ of the cluster lifetime $\tau_j$ distribution of the asset $i$  results:
\begin{equation}
\label{eq:entropy}
S_i(\tau_j,n) = S_0 + \log \tau_j^D + \frac{\tau_j}{n} \quad ,
\end{equation}
where $S_0$ is a constant, $\log \tau_j^D$ and $\tau_j/n$ respectively arises from power-law and exponentially correlated cluster duration. The subscript $j$ refers to the single cluster duration and will be suppressed in the forthcoming discussion for simplicity.
\par
The cluster entropy index $I_i (n) $ of a relevant random variable, e.g. the return of a given asset $i$, can be described as: 
\begin{equation}
\label{eq:index}
I_i(n) = \sum_{\tau=1}^{m} S_i(\tau, n) + \sum_{\tau=m}^{N} S_i(\tau, n)\quad.
\end{equation}
The first sum is referred to the power law regime of the cluster duration probability distribution. The second sum is referred to the linear regime of the cluster duration probability distribution, i.e. the excess entropy term with respect to the logarithmic one. The index $m$ represents the threshold value of the cluster lifetime between the regimes.
%%%%%%%% RESULTS %%%%%%%%
\section{Results} As recalled in the Introduction, the cluster entropy and related  portfolio's weights have been estimated over a single period (i.e. a single temporal horizon of about six years) in  \cite{ponta2018information}.
In this work, the entropy ability to quantify dynamics and heterogeneity is exploited to estimate the weights of a \textit{multi-period portfolio}.  Such a construction is possible as the cluster entropy  estimates involve horizon dependence. The values of the cluster entropy weights can be directly compared to the results obtained by using the traditional mean-variance approach and Sharpe ratio maximization  shown in fig. \ref{fig:barsharpeWeights} and fig. \ref{fig:sharpeWeights}.
\begin{figure*}[h]
 \centering
 \label{fig:entropy}
 \includegraphics[width=0.32 \textwidth]{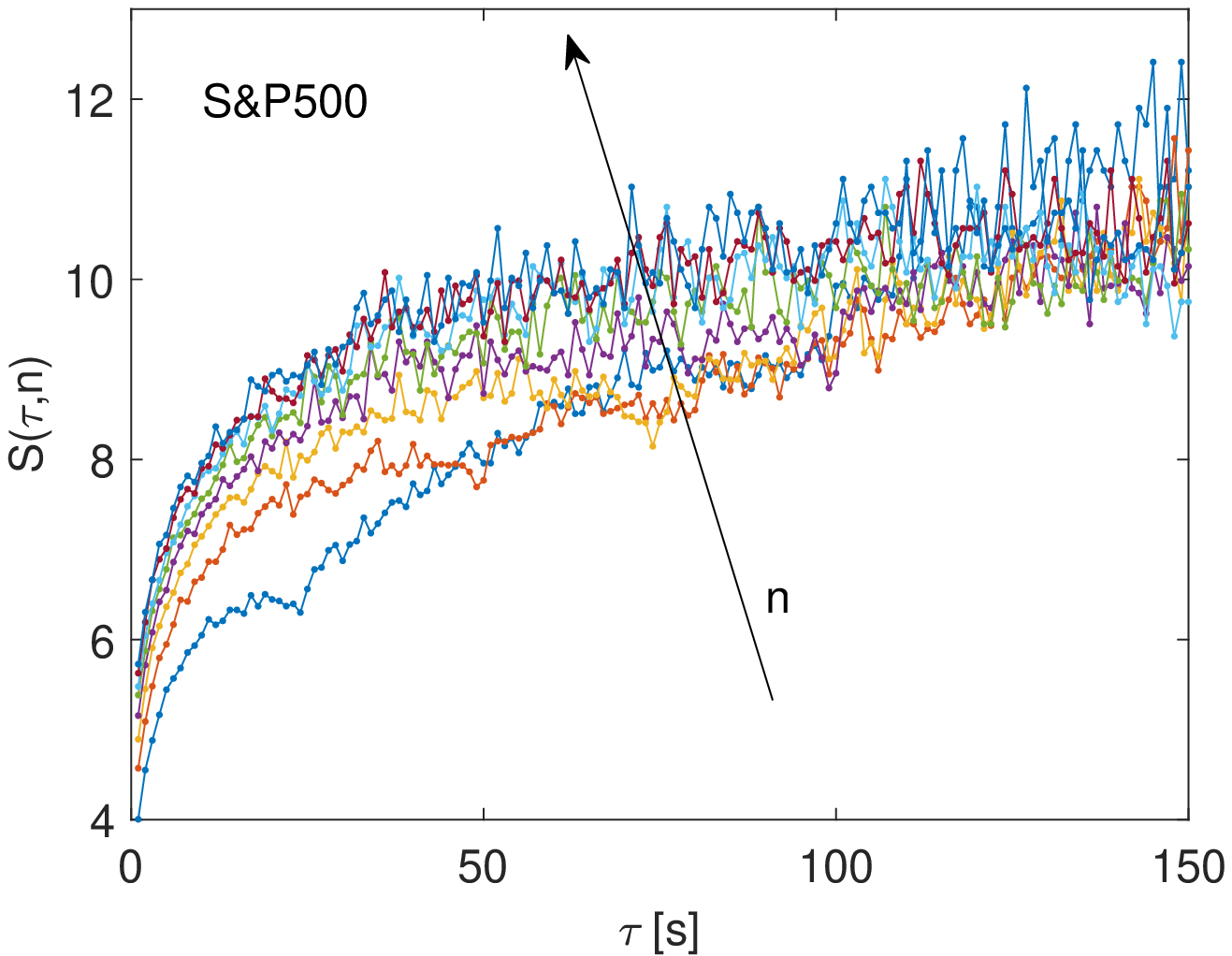}
 \includegraphics[width=0.32 \textwidth]{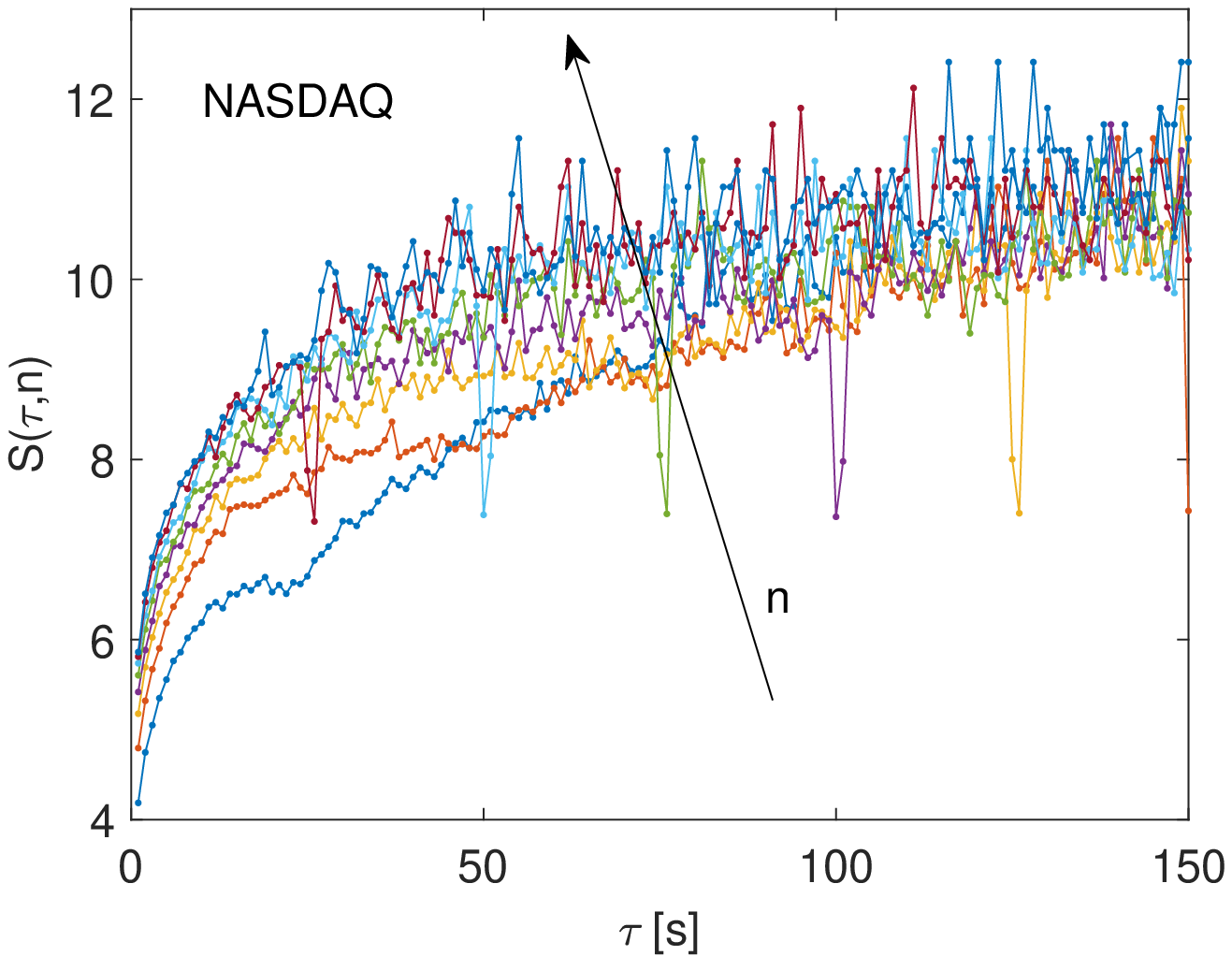}
\includegraphics[width=0.32 \textwidth]{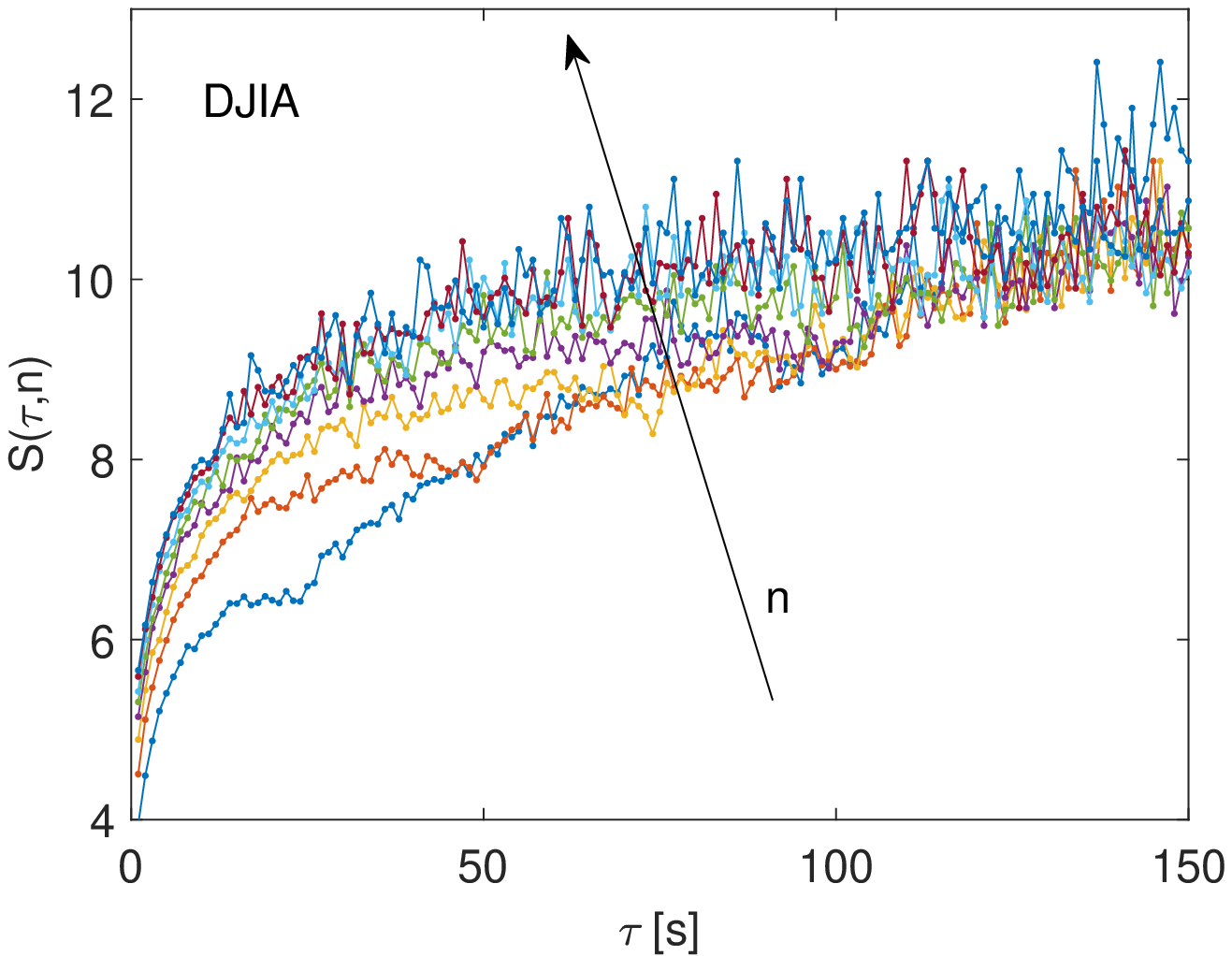}
 \includegraphics[width=0.32 \textwidth]{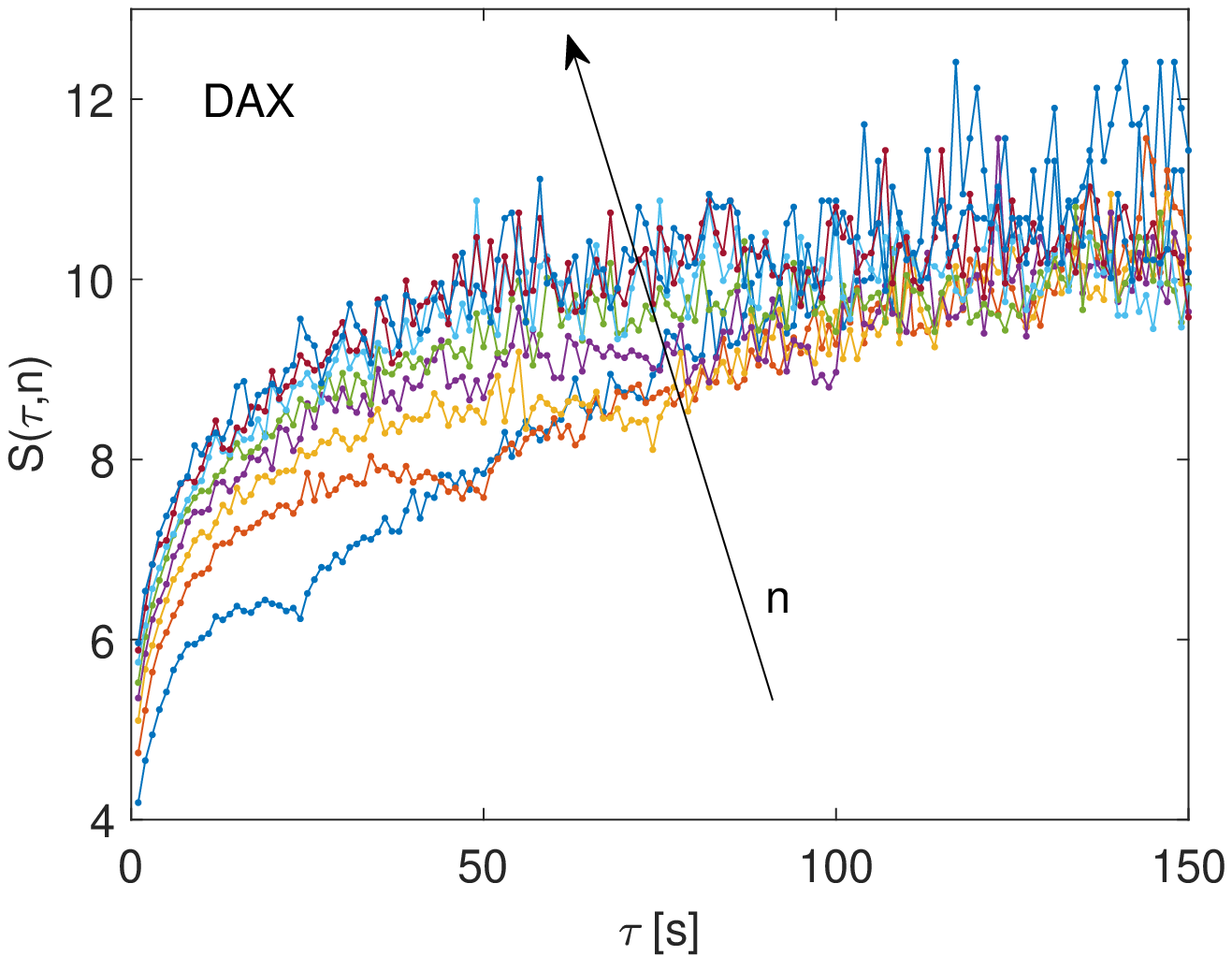}
 \includegraphics[width=0.32 \textwidth]{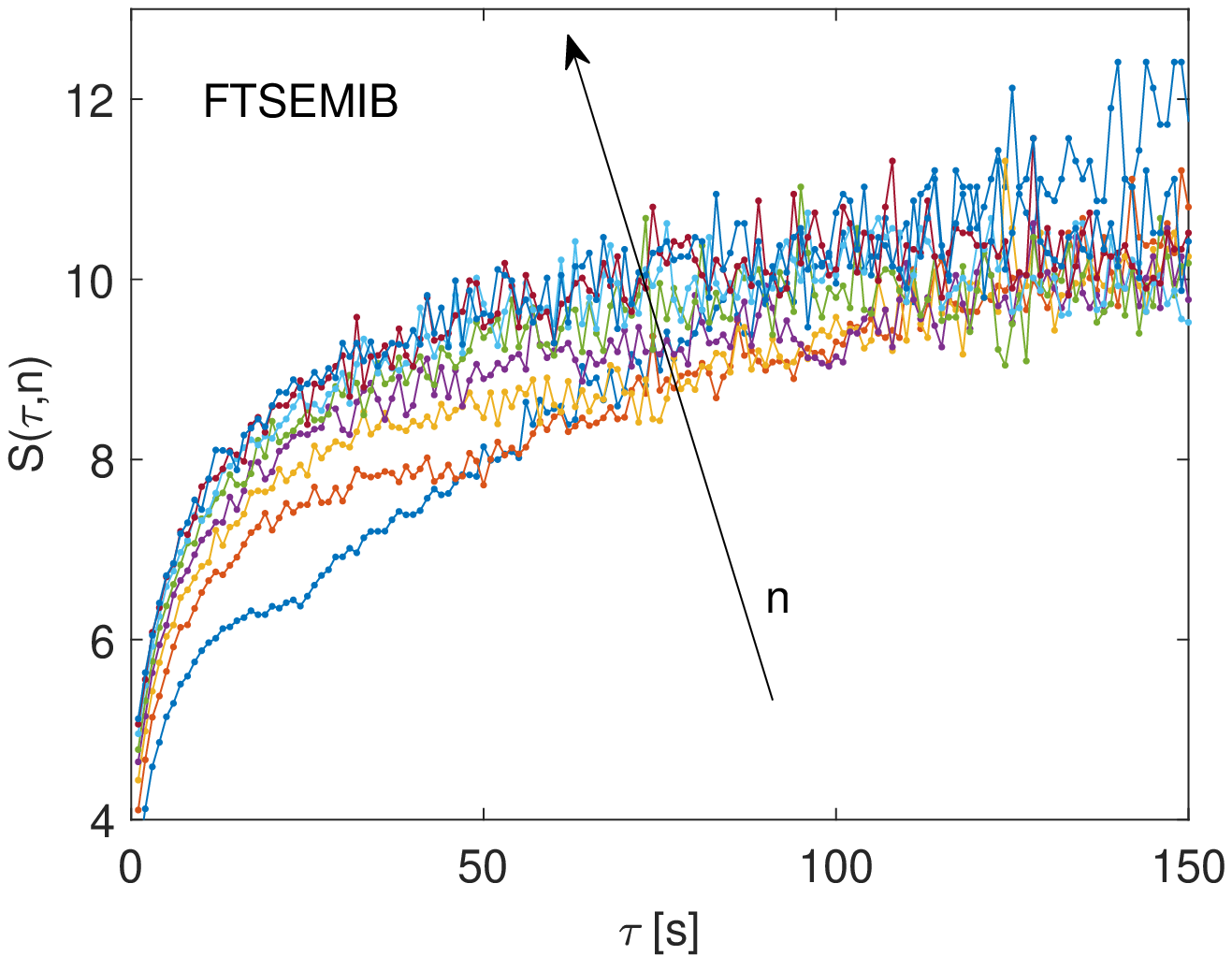}
\caption{\label{fig:entropy} Cluster entropy $S(\tau,n)$  calculated according to eq.~(\ref{eq:Shannonentropy}) for the probability distribution function in eq.~(\ref{eq:probdistr}) of the volatility series  of the linear return of tick-by-tick data of the S\&P500, NASDAQ, DJIA, 
DAX and FTSEMIB assets (further details in Table \ref{tab:data}). Time series have same length $N=492023$. The volatility is calculated according to eq.~(\ref{eq:volatility}) with window $T = 180s$ for all the five graphs. The plots refer to the horizon ${\cal{M}}=12$, i.e twelve monthly periods sampled out of the  year 2018. The different plots refer to different values of the moving average
window $n$ (here $n$ ranges from  $25s$ to  $200s$ with step $25s$).  Further results of the cluster entropy for a broad set of relevant parameters (volatility window $T$ and temporal horizon $\cal{M}$) can be found in the figures included in the Supplementary Material attached to this letter.
%Ret Lin M=12 T=180
}
\end{figure*}
\par
The construction of the \textit{multi-period portfolio} is carried on by implementing the  procedure on five market time series. The datasets include tick-by-tick prices $y_t$  from  Jan. 1 to Dec. 31, 2018 downloaded from the Bloomberg terminal (further details provided in Table \ref{tab:data}).
Given the price series $y_t$ and the related time series of the returns $r_t$, the volatility is defined as:
\begin{equation} \label{eq:volatility}
    \sigma_{t,T} = \sqrt{\frac{\sum_{t=k}^{k+T}{(r_t-\mu_{t,T})^2}}{T- 1} } \hspace{15 pt},
\end{equation}
with the volatility window ranging from $k$ to $k+T$ and $\mu_{t,T}$ the expected returns over the window $T$ defined as:
\begin{equation} \label{eq:return}
\mu_{t,T} = \frac{1}{T} \sum_{t=k}^{k+T} r_t \hspace{15 pt}.
\end{equation}
\par
The cluster entropy $S_i(\tau_j,n)$ of the volatility series  is estimated  by introducing eq.~(\ref{eq:probdistr}) into eq.~(\ref{eq:Shannonentropy}) for different horizons $\cal{M}$ (twelve monthly horizons out of one year have been considered). Results obtained on volatility series (assets described in  Table~\ref{tab:data}) are plotted in fig.~\ref{fig:entropy}. 
The behaviour is consistent with the expectations provided by eq.~(\ref{eq:entropy}): a logarithmic trend is observed at small cluster lengths $\tau$, whereas a linear trend appears at larger $\tau$ values.
\par
The current approach takes a different perspective compared to the traditional portfolio strategies. The cluster entropy  is straightaway estimated from the financial market data with no assumption about the return distribution.
To obtain the portfolio weights, the cluster entropy index $I_{i}(n)$ of the volatility of each market $i$ is estimated by using  eq.~(\ref{eq:index}). Then the average index $I_i$ is calculated over the set of moving average values $n$:
\begin{equation}
\label{eq:avgindex}
I_i=\sum_{n} I_{i}(n) \quad . \end{equation}
The quantity  $I_{i}$  is a cumulative figure of diversity, a number suitable to quantify and compare information content in different markets. 
For each market $i$ and volatility windows $T$ the index $I_{i}$ estimated according to eq.~(\ref{eq:avgindex}) is normalized as follows:  
\begin{equation}
\label{eq:normalizedclusterindex}
{w}_{i,\mathcal{C}}=
\frac{I_{i}}{  \sum_{i=1}^{\cal{N}_{\cal{A}}} I_{i}   } \quad ,
\end{equation}
to satisfy the condition $\sum_{i=1}^{\cal{N}_{\cal{A}}}{w}_{i,\mathcal{C}}=1$. 
The quantities ${w}_{i,\mathcal{C}}$ build the portfolio according to the  probability of the riskiest assets in the case of high-risk propensity of the investor.  Alternative estimates based on the cluster entropy index $I_i$ might be easily carried on for low-risk  profiles.
\par
\begin{figure*}[]
 \centering
 \includegraphics[width=0.32 \textwidth]{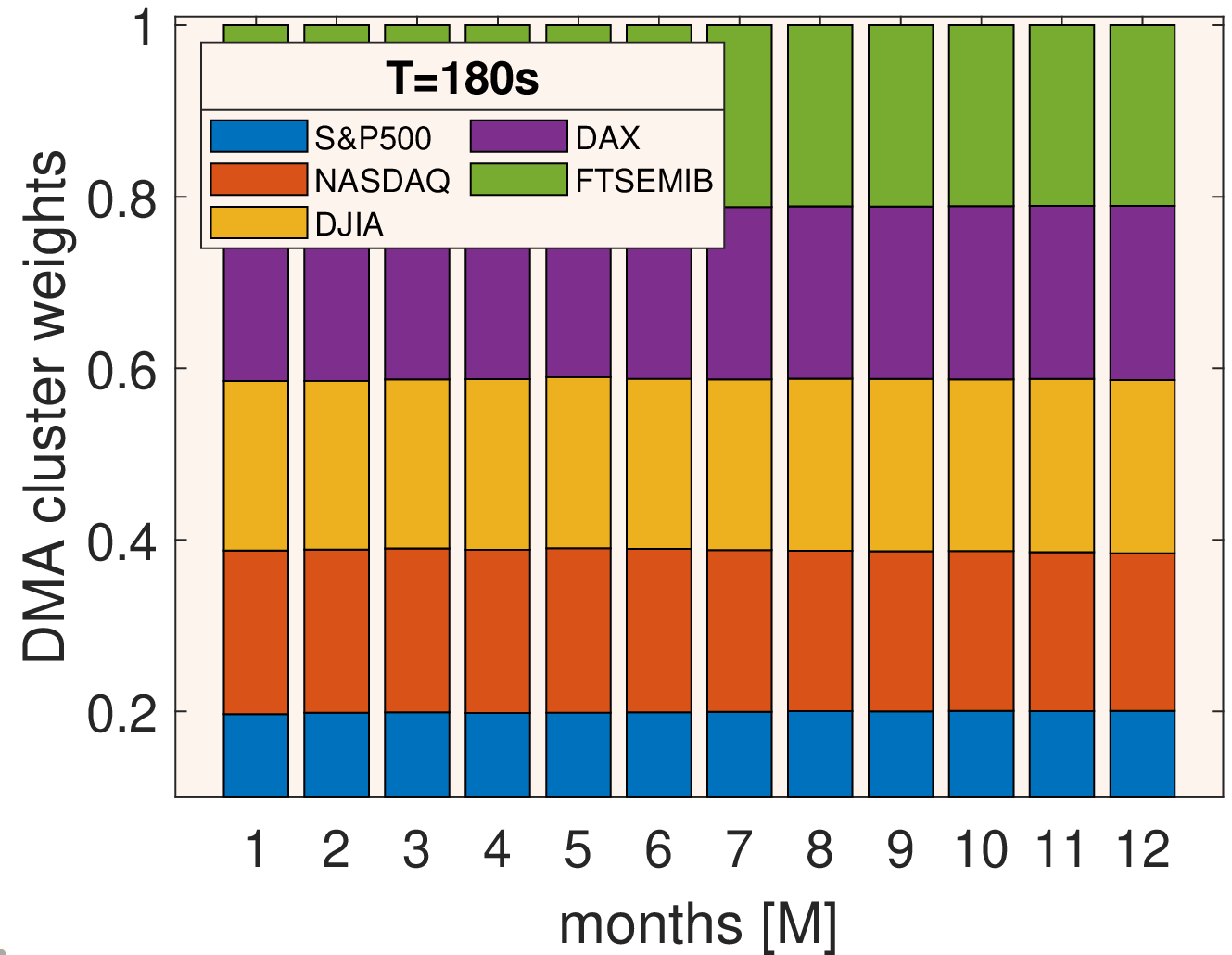}
 \includegraphics[width=0.32 \textwidth]{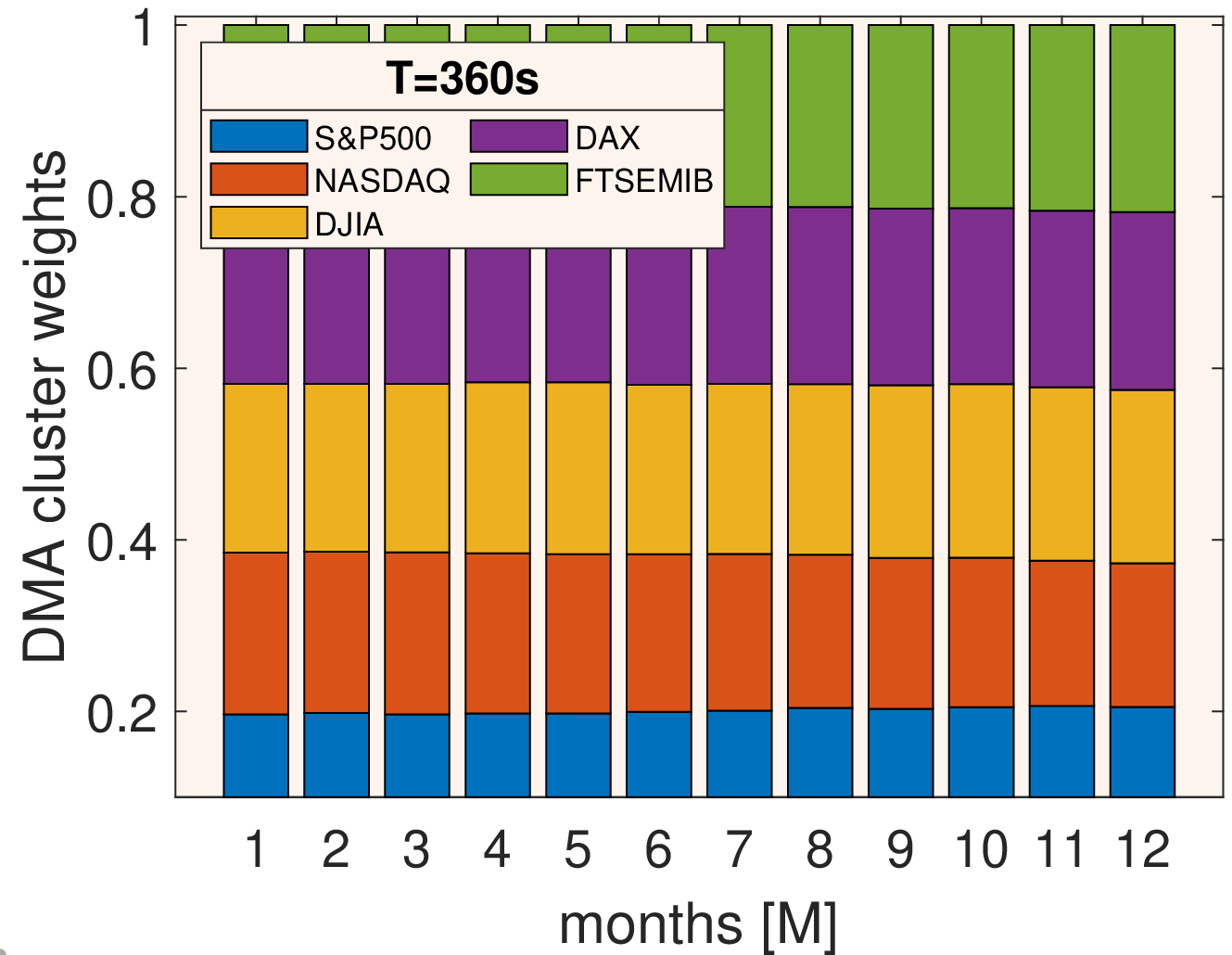}
 \includegraphics[width=0.32 \textwidth]{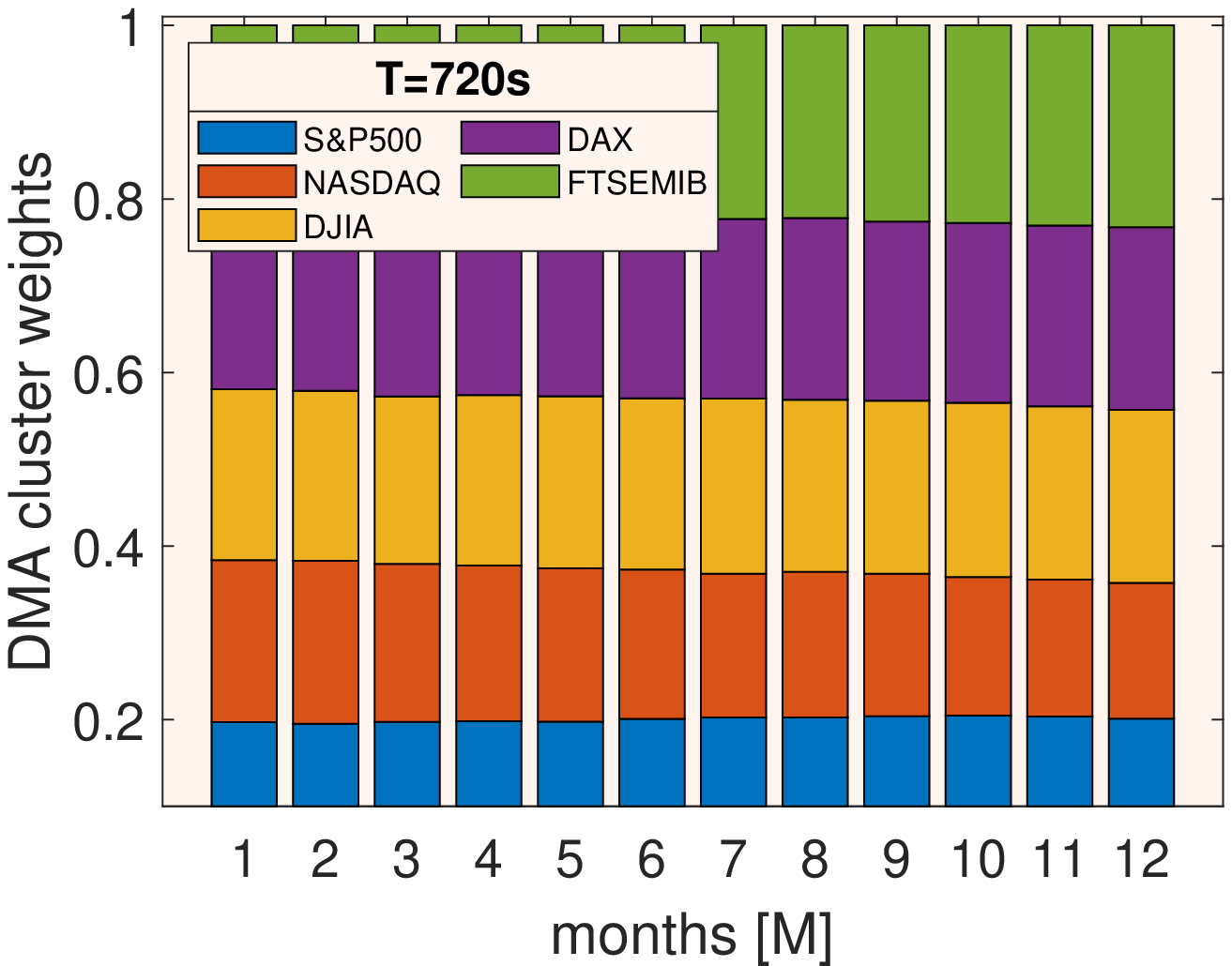}
 \caption{\label{fig:bardmaweights}  Portfolio weights $w_{i,\cal{C}}$ vs. investment horizon $\cal{M}$. The weights have been calculated according to the detrended cluster entropy approach 
 eqs.~(\ref{eq:index},\ref{eq:normalizedclusterindex}) by considering the case of high-risk propensity i.e. by maximizing  variance/volatility. The cluster entropy refers to the volatility series estimated according to eq.~(\ref{eq:volatility}) respectively with window $T=180s$, $T=360s$ and $T=720s$. At small horizons $\cal{M}$ and small volatility windows $T$ the weights take values close to those expected according to the  $1/\cal{N}_{\cal{A}}$ uniform distribution. This behaviour is related to the high unpredictability of the price and the associated risk (volatility) at short $\cal{M}$.  As horizon $\cal{M}$ and volatility window $T$ increase, a less diversified set of values is obtained. The decreased level of diversity is  consistent with the increased amount of information gathered along the series as time advances. (Further data of the portfolio weights  can be found in the Supplementary Material attached to this letter). 
}
 \end{figure*}
\begin{figure*}[h]
 \centering
 \includegraphics[width=0.32 \textwidth]{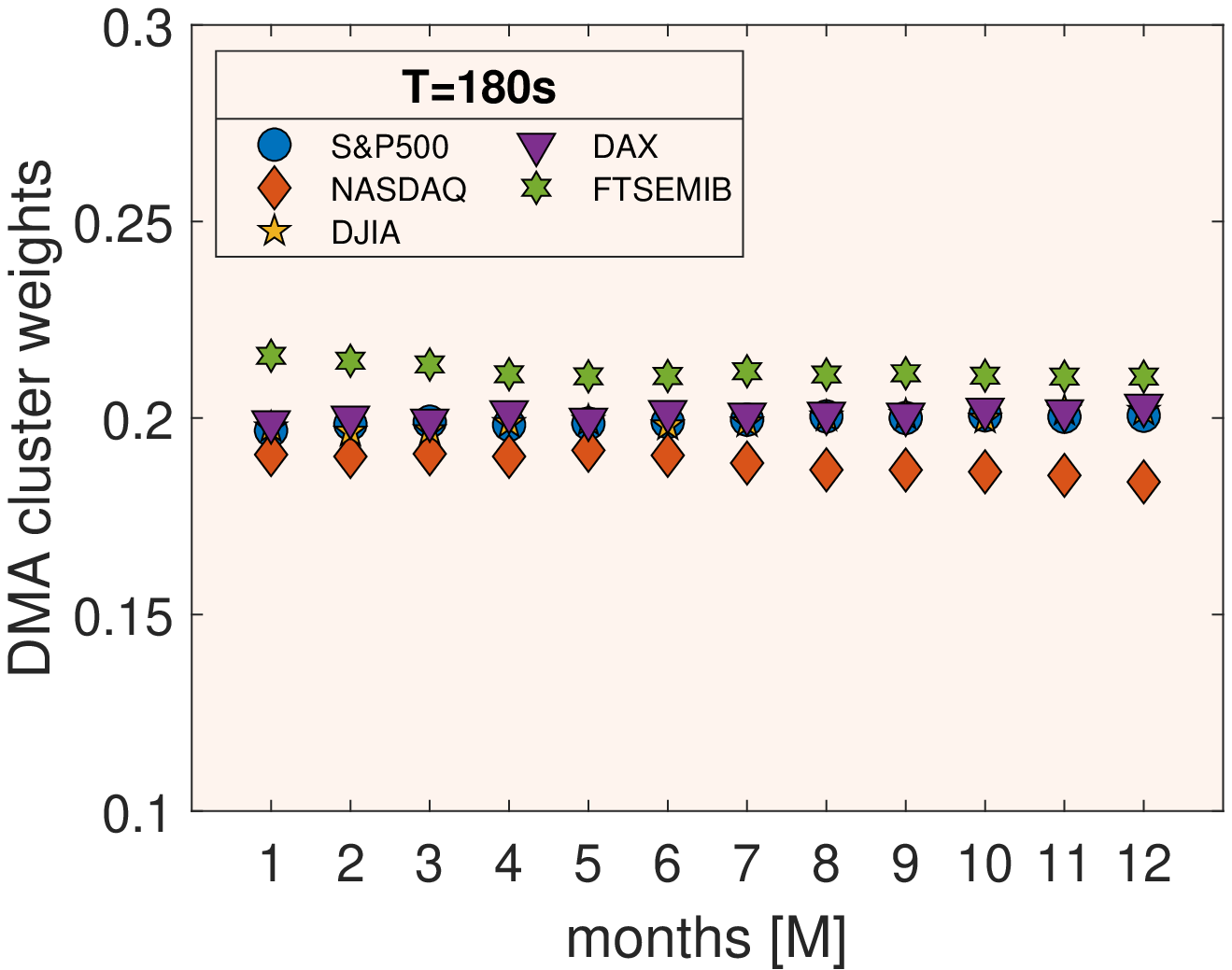}
 \includegraphics[width=0.32 \textwidth]{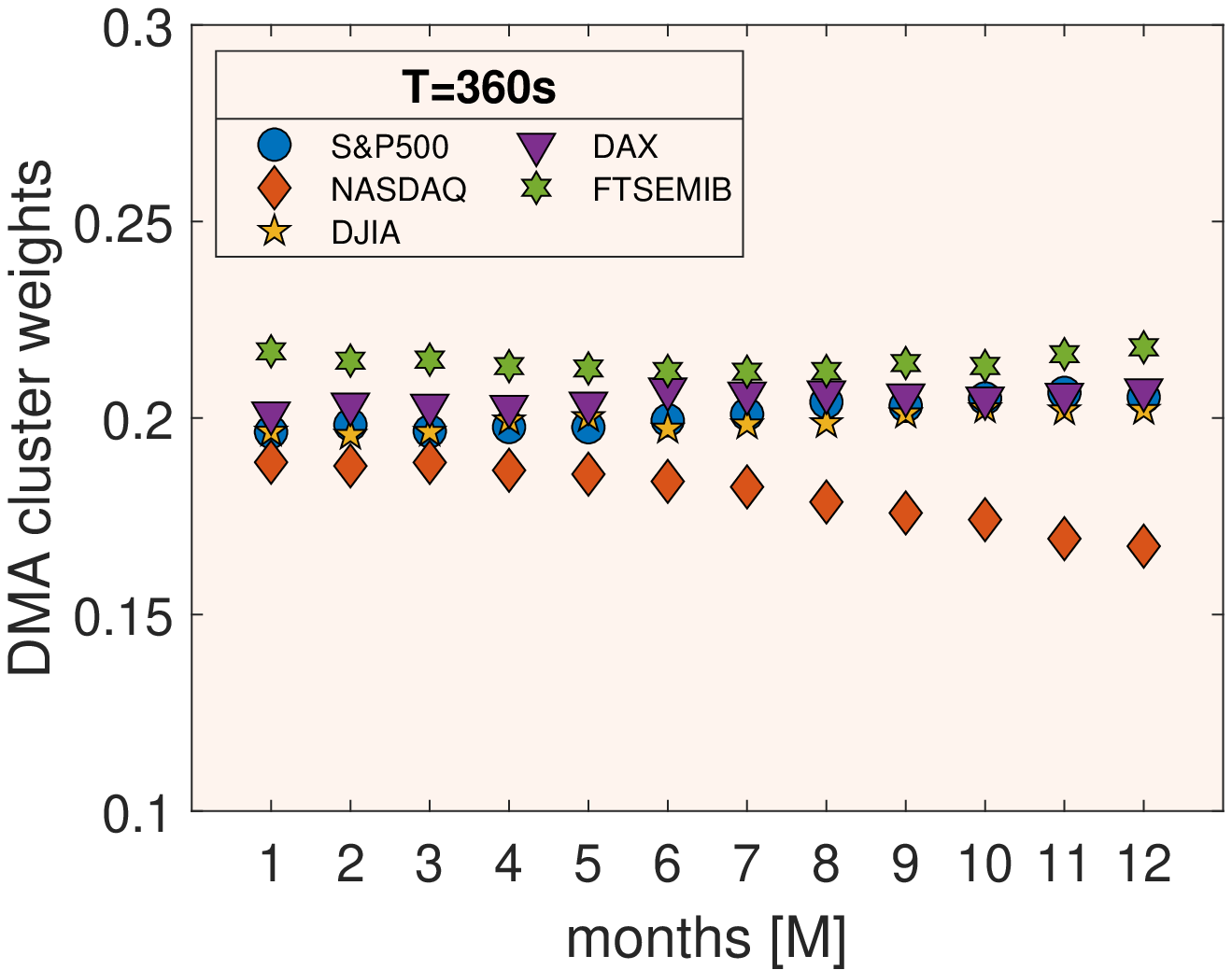}
 \includegraphics[width=0.32 \textwidth]{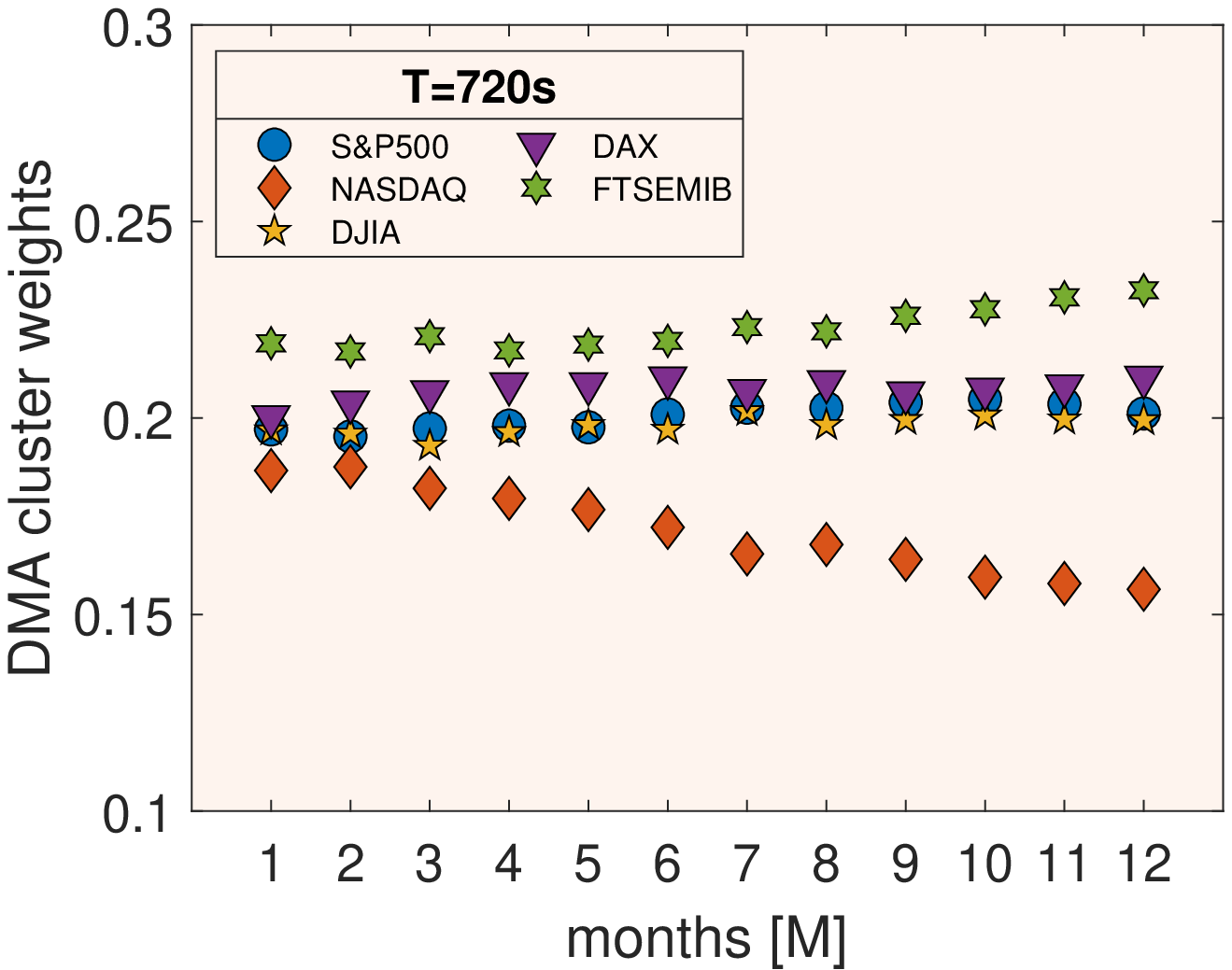}
 \caption{\label{fig:dmaweigths} Same as fig.~\ref{fig:bardmaweights} but with scattered plots.}
\end{figure*}

The weights  ${w}_{i,\mathcal{C}}$ are plotted in fig.~\ref{fig:bardmaweights} and fig.~\ref{fig:dmaweigths} for the five assets. At short horizons $\cal{M}$ and small volatility windows $T$, the weights take values close to $0.2$ as it would be expected by a uniform wealth allocation. The weights distribution, with  values close to $1/\cal{N}_{\cal{A}}$, is related to the low predictability degree of  price series and associated risk (volatility) given the limited amount of data at short periods and volatility windows.  As  $\cal{M}$ and $T$ increase, a less diversified weights distribution emerges consistently with the increased amount of information gathered along the widened temporal horizon.   Further results of the portfolio weights according to cluster entropy model can be found in the Supplementary Material attached to this letter. 
%%%%%%CONCLUSIONS %%%%%%%%%%
\section{Conclusion}
An innovative, dynamic and robust perspective of investment, based on the detrending moving average cluster entropy approach, is offered by a multi-period portfolio strategy that does not rely on the flawed assumptions of normal and stationary distribution of market data.  
The strategy is implemented on volatility of high-frequency market series (details in Table \ref{tab:data}) over multiple consecutive horizons $\cal{M}$. The high degree of stability, diversity and reliability of the portfolio weights can be indeed appreciated by the results shown in fig.~\ref{fig:bardmaweights} and in fig.~\ref{fig:dmaweigths}. A continuous set of values of portfolio weights with a smooth and sound dependence on the horizon can be observed.   
The entropy-based estimate of the portfolio is straightaway obtained from the stationary detrended distribution of the financial series rather than by using the unrealistic mean-variance hypothesis of Gaussian returns.  
At short volatility windows (e.g. $T=180s$ in fig.~\ref{fig:dmaweigths}),  weights take values close to the equally distributed $1/\cal{N}_{\cal{A}}$ portfolio.  These results are consistent with expected investment strategies where volatility (risk) does not give a prominent contribution. 
As $T$ increases, the volatility plays a relevant role in the weights estimate which deviates from the uniform  distribution.
\par
The proposed approach  uses a stationary set of variables (i.e. the detrended cluster durations $\tau_j$ of the return and volatility series rather than non-stationary and not-normal variables as asset returns and volatility) \cite{carbone2004analysis,carbone2007scaling,carbone2013information,ponta2018information,murialdo2020long,ponta2021information}.  The basic drawbacks of the traditional mean-variance approach are thus removed at their roots.
Clustering methods have demonstrated ability to obtain sound analysis of data with applications in various fields 
 including  portfolio strategies \cite{duran2013cluster,aghabozorgi2015time,iorio2018p,puerto2020clustering,tayali2020novel,tola2008cluster,massahi2020development}. Our approach is based on the joint adoption of clustering and information measure. 
Several developments can be envisaged, as for example cluster entropy portfolio optimization based on the cross-correlation cluster distance measures and Kullback-Leibler entropy.

\acknowledgments{
P.M. acknowledges financial support from FuturICT 2.0 a FLAG-ERA Initiative within the
Joint Transnational Calls 2016, Grant Number: JTC-2016-004}
%\bibliographystyle{eplbib.bst}
%\bibliography{Information2020,portfolios}

\end{document}